\documentclass{ws-ijmpa}
\usepackage[super,compress]{cite}
\usepackage{graphicx}
\begin{document}

\markboth{G. L.~Klimchitskaya, U.~Mohideen \& V.~M.~Mostepanenko}
{The Casimir effect in graphene systems: Experiment and theory}

\catchline{}{}{}{}{}

\title{The Casimir effect in graphene systems: Experiment and theory }

\author{{G.~L.~Klimchitskaya,${}^{1,2}$  U.~Mohideen${}^{3}$
{{and}}
 V.~M.~Mostepanenko${}^{1,2,4}$}}

\address{${}^1$Central Astronomical Observatory at Pulkovo of the
Russian Academy of Sciences, \\Saint Petersburg,
196140, Russia\\
${}^2$Peter the Great Saint Petersburg
Polytechnic University, \\Saint Petersburg, 195251, Russia\\
${}^3$Department of Physics and Astronomy, University of California, \\
Riverside, California 92521, USA\\
${}^4$Kazan Federal University, Kazan, 420008, Russia\\
g\_klimchitskaya@mail.ru, Umar.Mohideen@ucr.edu, vmostepa@gmail.com}

\maketitle

\begin{history}
\received{Day Month Year}
\revised{Day Month Year}
\end{history}

\begin{abstract}
The Casimir effect in graphene systems is reviewed with emphasis made on the 
large thermal correction to the Casimir force predicted at short separations 
between the test bodies. The computational results for the Casimir pressure 
and for the thermal correction are presented for both pristine graphene and 
real graphene sheets, which possess nonzero energy gap and chemical potential, 
obtained by means of exact polarization tensor. Two experiments on measuring 
the gradient of the Casimir force between an Au-coated sphere and graphene-
coated substrates performed by using a modified atomic force microscope 
cantilever-based technique are described. It is shown that the measurement
data of both experiments are in agreement with theoretical predictions of 
the Lifshitz theory using the polarization tensor. Additionally, several 
important improvements made in the second experiment, allowed to demonstrate 
the predicted large thermal effect in the Casimir interaction at short 
separations. Possible implications of this result to resolution of 
long-term problems of Casimir physics are discussed.
\end{abstract}

\ccode{PACS Nos.: 68.65.Pq; 12.20.Fv; 12.20.Ds; 12.20.-m }

\section{Introduction}
\label{secKMM:1}
\newcommand{\kb}{{k_{\bot}}}
\newcommand{\skb}{{k_{\bot}^2}}
\newcommand{\vk}{{\mbox{\boldmath$k$}}}
\newcommand{\rv}{{\mbox{\boldmath$r$}}}
\newcommand{\ve}{{\varepsilon}}
\newcommand{\okb}{{(\omega,k_{\bot})}}
\newcommand{\xkb}{{(i\xi_l,k_{\bot})}}

The Casimir effect \cite{KMM1} is an extraordinary physical phenomenon which has been
much investigated in several fields of physics for almost 75 years after its discovery.
According to Casimir's result, two parallel ideal metal planes in vacuum attract each
other by a force which depends only on the Planck constant $\hbar$, speed of light $c$,
and the separation between the planes. This force originates from the zero-point
fluctuations of the
quantum electromagnetic field. If the planes are at temperature $T$ in thermal equilibrium
with the environment, the force value is also determined by the value of $T$ and the
Boltzmann constant $k_B$. The thermal Casimir force is caused by the joint action of
the zero-point and thermal fluctuations.

Lifshitz\cite{KMM2,KMM2a} developed a general theory of the Casimir force between two material
plates described by the frequency-dependent dielectric permittivities. In the last
20 years this theory has been shown to have problems when comparing theoretical
predictions with the
measurement data of high precision experiments and with the requirements of thermodynamics
(see an extensive review\cite{KMM3,KMM4,KMM5,KMM6} and the most recent
experiments\cite{KMM7,KMM8,KMM9,KMM10}). One conceivable reason for these problems is
that the dielectric permittivities describing the response of the plate materials to
electromagnetic fluctuations are to a large measure of phenomenological character and
may fail to account for all subtle features of the field-matter interaction.

In this regard graphene, which is a 2D sheet of carbon atoms packed in a hexagonal
lattice\cite{KMM11,KMM11a,KMM11b,KMM12,KMM13} is of immediate interest to theorists and
experimentalists working in the Casimir effect. The reason is that at energies below
some definite value (which was estimated\cite{KMM14} as approximately equal to 3~eV)
graphene is well described in the framework of the Dirac model as a set of either
massless or very light electronic quasiparticles. The quantum field of these
quasiparticles satisfies the relativistic Dirac equation in 2+1
dimensions\cite{KMM11,KMM11a,KMM11b,KMM12,KMM13} with the only difference
that the speed of light
is replaced with the Fermi velocity $v_{\rm F}\approx c/300$.

Given that graphene is so simple a physical system, it becomes possible
to find its dielectric response to electromagnetic fluctuations starting from the
first principles of quantum electrodynamics at nonzero temperature without 
resorting
to phenomenological methods. This was done by finding the polarization tensor of
graphene\cite{KMM15,KMM16,KMM17,KMM18} which is equivalent to two spatially
nonlocal dielectric permittivities, the longitudinal one and the transverse one,
depending on both the frequency and the wave vector.\cite{KMM18a}
In so doing the Casimir
interaction between two graphene sheets is described by the standard Lifshitz
formulas where the reflection coefficients take the non-Fresnel form and are
expressed via the components of the polarization tensor. This offers strong
possibilities of reliable theoretical predictions for the Casimir interaction
between two graphene sheets and graphene-coated substrates which can be tested
experimentally and checked for a consistency with the requirements of
thermodynamics.

The first experiment on measuring the gradient of the Casimir force between an
Au-coated sphere and a graphene sheet on a Si-SiO${}_2$ substrate was performed
by means of an atomic force microscope (AFM) operated in the frequency shift
technique.\cite{KMM19} The measurement results were found to be in good agreement
with theoretical predictions by using the polarization tensor.\cite{KMM20}
Due to the small thickness of the SiO${}_2$ substrate,\cite{KMM21}
it was, however, not possible to detect the large thermal effect in the Casimir force
at short separations predicted previously for graphene.\cite{KMM22}

In the second experiment  on measuring the gradient of the Casimir force from
graphene by means of an AFM, a thicker SiO${}_2$ substrate was used. In addition,
the energy gap $\Delta$ caused by a nonzero mass of quasiparticles and the chemical
potential $\mu$ caused by the presence of impurities in the graphene sample
have been found utilizing scanning tunneling spectroscopy and Raman spectroscopy,
respectively.\cite{KMM23,KMM24} The measurement data were compared with theoretical
predictions using the polarization tensor and the existence of large thermal effect
was confirmed over the range of separations from 250 to 517~nm.

In this review, we consider all the above results with due regard to the investigation
of the Casimir effect in graphene systems performed using some other techniques.
In Sec.~\ref{secKMM:2}, the Lifshitz formulas describing the Casimir interaction
between two graphene sheets and between graphene-coated substrates using the formalism
of the polarization tensor are presented.  Section~\ref{secKMM:3} is devoted to the
thermal Casimir force from the sheets of pristine and real graphene.
In Sec.~\ref{secKMM:4}, the first experiment on measuring the gradient of the
Casimir force from graphene is described. Demonstration of the unusual thermal effect
in the Casimir interaction from graphene made in the second experiment is contained
in Sec.~\ref{secKMM:5}. Finally, in Sec.~\ref{secKMM:6}, the reader will find our
conclusions and discussion of the obtained results for the Casimir force from
graphene with their possible implication to other materials.

\section{Lifshitz Formula, Electromagnetic Response of Graphene, and
the Polarization Tensor}
\label{secKMM:2}

According to the scattering theory approach to electrodynamic Casimir
forces,\cite{KMM24a,KMM24b,KMM25}
the standard Lifshitz formulas for the Casimir free energy per unit area  ${\cal F}$ and
pressure $P$, originally derived \cite{KMM2} for the case of two semispaces, remain valid
for any two planar structures with appropriately defined reflection coefficients
$R_{\rm TM}^{(n)}$ and $R_{\rm TE}^{(n)}$, $n=1,\,2$, for the transverse magnetic and
transverse electric polarizations of the electromagnetic field. They are as follows:
\begin{equation}
{\cal F}(a,T)=\frac{k_BT}{2\pi}\sum_{l=0}^{\infty}{\vphantom{\sum}}^{\prime}
\int_{0}^{\infty}\!\!\!\kb d\kb\sum_{\lambda}\ln\left[1-
R_{\lambda}^{(1)}\xkb R_{\lambda}^{(2)}\xkb e^{-2aq_l}\right]
\label{eqKMM1}
\end{equation}
\noindent
and
\begin{equation}
{P}(a,T)=-\frac{k_BT}{\pi}\sum_{l=0}^{\infty}{\vphantom{\sum}}^{\prime}
\int_{0}^{\infty}\!\!\!q_l\kb d\kb\sum_{\lambda}\left[
\frac{e^{2aq_l}}{R_{\lambda}^{(1)}\xkb R_{\lambda}^{(2)}\xkb}-1\right]^{-1}\!\!\! .
\label{eqKMM1a}
\end{equation}
\noindent
Here, $a$ is a separation between the plane structures, the prime on the summation in $l$
divides the term with $l=0$ by 2, $\kb$ is the magnitude of the wave vector projection on
the planar structures, $\xi_l=2\pi k_BTl/\hbar$ are the Matsubara frequencies,
$q_l=(\skb+\xi_l^2/c^2)^{1/2}$, and $\lambda=({\rm TM,\,TE})$.
The quantities $R_{\lambda}^{(n)}$ may have a meaning of the reflection coefficients
on metallic or dielectric plates, on graphene sheets or on the graphene-coated substrates.

There are many theoretical approaches to calculation of the reflection coefficients on
a graphene sheet based on a hydrodynamic model,\cite{KMM26,KMM27,KMM28} Kubo
theory,\cite{KMM29,KMM30,KMM31} density-density correlation functions found in the
random phase approximation,\cite{KMM22,KMM32,KMM33,KMM34,KMM35,KMM36} in-plane and
out-of-plane electrical conductivities of graphene obtained by means of the 2D Drude
model, Kubo formula etc.\cite{KMM31,KMM34,KMM36,KMM36a,KMM37,KMM37a,KMM37b}
It should be noted, however, that the exact expressions for the correlation functions
and conductivities of graphene at any nonzero temperature remained unknown.
Specifically, the conductivities calculated using the Kubo formula include the
phenomenological relaxation parameter and neglect the energy gap of graphene.

As mentioned in Sec.~\ref{secKMM:1}, at energies below 3~eV the Dirac model provides
a comprehensive fundamental description of graphene (note that the first absorption peak
of graphene takes place at larger energy of $\hbar\omega\approx 4.59~$eV).
Taking into account that measurements of the Casimir interaction by means of the dynamic
AFM are performed at separations exceeding 200~nm, which correspond to the
characteristic energies $\hbar c/(2a)<0.5~$eV, an application of the Dirac model for
calculation of the Casimir force from graphene is fully justified.

In the framework of the Dirac model, the reflection coefficients on a graphene sheet
can be expressed via its polarization tensor
$\Pi_{\beta\gamma,l}=\Pi_{\beta\gamma}(i\xi_l,\kb,T,\Delta,\mu)$ found using the formalism
of quantum electrodynamics at nonzero temperature.\cite{KMM15,KMM16,KMM17,KMM18}
For real graphene sheets the components of the polarization tensor ($\beta,\gamma=0,\,1,\,2$)
depend on the energy gap $\Delta=2mv_{\rm F}^2$, where $m$ is the mass of quasiparticles,
and on the chemical potential $\mu$. For generality, we present the reflection coefficients on
thick material plates with the dielectric permittivities $\ve_l^{(n)}=\ve^{(n)}(i\xi_l)$
coated with a graphene sheet\cite{KMM20,KMM38,KMM39}
\begin{eqnarray}
&&
R_{\rm TM}^{(n)}(i\xi_l,k_{\bot})=
\frac{\hbar k_{\bot}^2[\varepsilon_l^{(n)}q_l-k_l^{(n)}]+
q_lk_l^{(n)}\Pi_{00,l}} {\hbar k_{\bot}^2[\varepsilon_l^{(n)}q_l+k_l^{(n)}]+
q_lk_l^{(n)}\Pi_{00,l}},
\nonumber\\
&&
R_{\rm TE}^{(n)}(i\xi_l,k_{\bot})=
\frac{\hbar k_{\bot}^2[q_l-k_l^{(n)}]-
\Pi_{l}} {\hbar k_{\bot}^2[q_l+k_l^{(n)}]+\Pi_{l}},
\label{eqKMM2}
\end{eqnarray}
\noindent
where $k_l^{(n)}=[\skb+\ve_l^{(n)}\xi_l^2/c^2]^{1/2}$ and the quantity $\Pi_l$ is expressed
via the trace of the polarization tensor and its 00-component according to:
\begin{equation}
\Pi_l=\skb\Pi_{\beta,l}^{\,\beta}-q_l^2\Pi_{00,l}.
\label{eqKMM3}
\end{equation}

The reflection coefficients on the freestanding (with no substrate) graphene sheet
are obtained from (\ref{eqKMM2}) by putting $\ve_l^{(n)}=1$, $k_l^{(n)}=q_l$.
The standard (Fresnel) reflection coefficients on the plates made of ordinary materials
are obtained from (\ref{eqKMM2}) by putting $\Pi_{00,l}=\Pi_l=0$.

It is convenient to present the quantities $\Pi_{00,l}$ and $\Pi_l$ in the form
\begin{equation}
\Pi_{00,l}=\Pi_{00,l}^{(0)}+\Pi_{00,l}^{(1)},
\quad
\Pi_{l}=\Pi_{l}^{(0)}+\Pi_{l}^{(1)},
\label{eqKMM4}
\end{equation}
\noindent
where $\Pi_{00,l}^{(0)}$ and $\Pi_{l}^{(0)}$ refer to the polarization tensor of graphene
with perfect hexagonal lattice with no impurities ($\mu=0$), zero temperature ($T=0$),
and any value of the energy gap $\Delta$. By construction, the terms
$\Pi_{00,l}^{(0)}$ and $\Pi_{l}^{(0)}$ in (\ref{eqKMM4}) do not depend on $T$ as a parameter
but only implicitly, through the Matsubara frequencies, as they are calculated at
$\omega=i\xi_l$. By contrast, the terms
$\Pi_{00,l}^{(1)}$ and $\Pi_{l}^{(1)}$ in (\ref{eqKMM4}) include an explicit dependence
of the polarization tensor on $T$ as a parameter. They also depend on the chemical
potential $\mu$ and the energy gap $\Delta$.

The terms $\Pi_{00,l}^{(0)}$ and $\Pi_{l}^{(0)}$ are given by\cite{KMM15,KMM16}
\begin{equation}
\Pi_{00,l}^{(0)}=\frac{\alpha\hbar k_{\bot}^2}{\tilde{q}_l}\Psi(D_l),
\quad
\Pi_{l}^{(0)}=\alpha\hbar k_{\bot}^2\tilde{q}_l\Psi(D_l),
\label{eqKMM5}
\end{equation}
\noindent
where $\alpha=e^2/(\hbar c)$ is the fine structure constant and the following notations
are introduced
\begin{equation}
\Psi(x)=2\left[x+(1-x^2)\arctan\frac{1}{x}\right],
\quad
\tilde{q}_l=\sqrt{\frac{v_F^2}{c^2}k_{\bot}^2+\frac{\xi_l^2}{c^2}},
\quad
D_l=\frac{\Delta}{\hbar c\tilde{q}_l}.
\label{eqKMM6}
\end{equation}

The second terms on the right-hand side of (\ref{eqKMM4}) take the form \cite{KMM18,KMM39}
\begin{eqnarray}
&&
\Pi_{00,l}^{(1)}=\frac{4\alpha\hbar c^2\tilde{q}_l}{v_F^2}
\int_{D_l}^{\infty}\!\!du \left(\sum_{\kappa=\pm 1}
\frac{1}{e^{B_lu+\kappa\frac{\mu}{k_BT}}+1}\right)
\nonumber \\
&&
~~~~~~~~\times
\left[1-{\rm Re}\frac{1-u^2+2i\gamma_lu}{\left(1-u^2
+2i\gamma_lu+D_l^2-\gamma_l^2D_l^2\right)^{1/2}}
\right],
\nonumber \\[1mm]
&&
\Pi_{l}^{(1)}=-\frac{4\alpha\hbar \tilde{q}_l\xi_l^2}{v_F^2}
\int_{D_l}^{\infty}\!\!du \left(\sum_{\kappa=\pm 1}
\frac{1}{e^{B_lu+\kappa\frac{\mu}{k_BT}}+1}\right)
\nonumber \\
&&
~~~~~~~~\times\left[1-{\rm Re}\frac{(1+i\gamma_l^{-1}u)^2+(\gamma_l^{-2}-1)
D_l^2}{\left(1-u^2
+2i\gamma_lu+D_l^2-\gamma_l^2D_l^2\right)^{1/2}}
\right],
\label{eqKMM7}
\end{eqnarray}
\noindent
where $\gamma_l\equiv\xi_l/(c\tilde{q}_l)$ and
$B_l\equiv\hbar c\tilde{q}_l/(2k_BT)$.

Note that the density-density correlation functions and the spatially nonlocal dielectric
permittivities of graphene are uniquely determined by its polarization tensor.\cite{KMM40}
For instance, the transverse and longitudinal permittivities are given by
\begin{equation}
\ve_l^{\rm Tr}=1+\frac{c^2}{2\hbar\kb\xi_l^2}\Pi_l \qquad
\ve_l^{\rm L}=1+\frac{1}{2\hbar\kb}\Pi_{00,l}.
\label{eqKMM8}
\end{equation}
\noindent
Thus, the exact expressions (\ref{eqKMM4})--(\ref{eqKMM7}) for the polarization tensor
also provide the respective expressions for the density-density correlation functions of
graphene at any temperature and make both formalisms equivalent.\cite{KMM40}

Using the Casimir free energy ${\cal F}$ (\ref{eqKMM1}) with the reflection coefficients
(\ref{eqKMM2})--(\ref{eqKMM7}), one can define the Casimir entropy
\begin{equation}
S(a,T)=-\frac{\partial{\cal F}(a,T)}{\partial T}
\label{eqKMM9}
\end{equation}
\noindent
and determine its behavior with vanishing temperature. For two pristine graphene sheets
($\ve_l^{(n)}=1$, $\Delta=\mu=0$) it was shown \cite{KMM41} that the Casimir entropy
goes to zero when the temperature vanishes in line with the Nernst heat theorem.
The same was proven \cite{KMM42} for two real graphene sheets characterized by nonzero
values of the energy gap and chemical potential. Similar results were
obtained \cite{KMM43,KMM44} for the Casimir-Polder entropy related to an atom interacting
with a graphene sheet (see also the review in Ref.~\citen{KMM45}).

One can conclude that the Lifshitz theory of the Casimir interaction between graphene
sheets is in perfect agreement with the requirements of thermodynamics if the dielectric
response of graphene is described by the formalism of the polarization tensor.

\section{Thermal Casimir Force from Sheets of Pristine and Real Graphene}
\label{secKMM:3}

The formalism of Sec.~\ref{secKMM:2} allows calculation of the thermal Casimir force
between pristine and real graphene sheets as well as between
different material plates and a graphene
sheet either freestanding or deposited on a substrate. Below we present several
characteristic results of this kind which give an idea of the thermal effect in the Casimir
force from graphene as opposed to ordinary materials.

\begin{figure}[b]
\vspace*{-8cm}
\centerline{\hspace*{-1.6cm}
\includegraphics[width=6.50in]{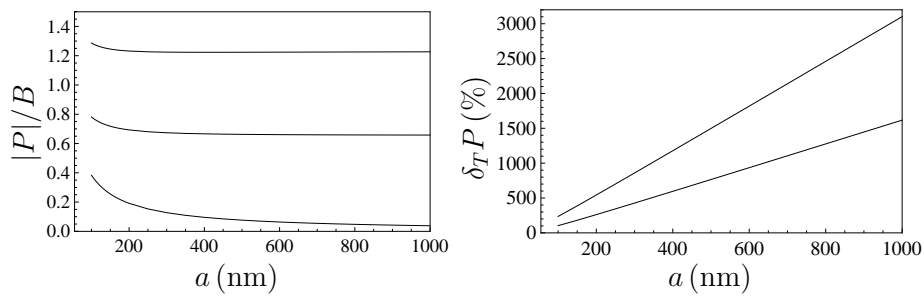}}
\vspace*{-11.2cm}
\caption{The normalized magnitudes of the Casimir pressure between two
pristine freestanding graphene sheets are shown as functions of separation.
The top and middle lines are computed at $T=300~$K using the full polarization tensor
and its zero-temperature contribution taken at the Matsubara frequencies, respectively.
The bottom line is computed at $T=0~$K (left).
The relative thermal corrections to  the Casimir pressure between two
pristine freestanding graphene sheets are shown as functions of separation by the
top and bottom lines computed at $T=300~$K using the full polarization tensor
and its zero-temperature contribution taken at the Matsubara frequencies,
respectively (right).
\protect\label{figKMM:1}}
\end{figure}
We start with the simplest configuration of two parallel sheets of pristine graphene at
room temperature $T=300~$K. The thermal Casimir pressure in this configuration is
computed\cite{KMM38} by (\ref{eqKMM1a}) and (\ref{eqKMM2})--(\ref{eqKMM7})
where one should put $\ve_l^{(n)}=1$ and $\Delta=\mu=0$. In Fig.~\ref{figKMM:1} (left),
the computational results for the magnitude of the Casimir pressure $|P|$ normalized to the
quantity $B=k_BT/(8\pi a^3)$ are shown as a function of separation by the top line.

To gain a better understanding of the role of thermal effects, we consider separately
the impact of the explicit and implicit thermal dependence on the Casimir pressure.
For this purpose, we repeat the same computations as above, but with only the first term of
the polarization tensor in (\ref{eqKMM4}) given by (\ref{eqKMM5}). This term does not
include the explicit dependence of the polarization tensor on $T$ as a parameter and depends
on temperature only implicitly through the Matsubara frequencies. The obtained computational
results for $|P|/B$, as a function of separation, are shown by the middle line in
Fig.~\ref{figKMM:1} (left).

Next, the bottom line in Fig.~\ref{figKMM:1} (left) is computed by the same formulas as the
middle line, but at $T=0~$K in the strict sense. This means that the summation over the
Matsubara frequencies in (\ref{eqKMM1}) is replaced with an integration in  continuous
frequency according to
\begin{equation}
k_BT\sum_{l=0}^{\infty}{\vphantom{\sum}}^{\prime}\to\frac{\hbar}{2\pi}\int_0^{\infty}d\xi.
\label{eqKMM10}
\end{equation}

Thus, a difference between the middle and bottom lines in Fig.~\ref{figKMM:1} (left)
illustrates the role of an implicit thermal effect in the Casimir pressure between two
graphene sheets whereas a difference between the top and middle lines shows the
contribution from an explicit dependence of the polarization tensor on temperature as
a parameter. As is seen in Fig.~\ref{figKMM:1} (left), in the region of separations
considered, both thermal effects contribute to the Casimir pressure roughly
equally.

The role of thermal effects in the Casimir pressure between the sheets of pristine graphene
can be expressed quantitatively by the relative thermal correction
\begin{equation}
\delta_TP(a,T)=\frac{\Delta_TP(a,T)}{P(a,0)}=\frac{P(a,T)-P(a,0)}{P(a,0)}.
\label{eqKMM11}
\end{equation}

In Fig.~\ref{figKMM:1} (right), the computational results for $\delta_TP$ at $T=300~$K
are presented as functions of separation by the top and bottom lines computed with the full
polarization tensor (\ref{eqKMM4}),  (\ref{eqKMM5}), (\ref{eqKMM7}) and the polarization
tensor (\ref{eqKMM5}) defined at $T=0$, respectively.
Thus, the top line presents the total thermal effect whereas the bottom line --- only the
implicit one. In fact both these effects are unexpectedly large. For example, at 100, 200,
600, and 1000~nm the total thermal correction constitutes 236\%, 544\%, 1818\%, and  3101\%
of the zero-temperature Casimir pressure, respectively. The implicit thermal correction
constitutes 104\%, 262\%, 935\%, and 1617\% at the same respective separations.

An unusually large thermal effect in the Casimir pressure between two graphene sheets at
short separations was first predicted \cite{KMM22} by using the method of density-density
correlation functions in the random phase approximation. It can be considered as unusual
because for ordinary materials at separations below a micrometer the thermal effect does
not exceed a fraction of a percent (a few percent effect predicted for metals when using
an extrapolation of the optical data by means of the Drude model was experimentally
excluded\cite{KMM3,KMM4,KMM5,KMM6,KMM7,KMM8,KMM9,KMM10}).
So the large magnitude of the thermal effect for graphene can be explained
physically\cite{KMM22,KMM24} by the fact that, in addition to the standard effective
temperature $k_BT_{\rm eff}=\hbar c/(2a)$ inherent to all materials, graphene is also
characterized by much lower effective temperature
$k_B\tilde{T}_{\rm eff}=\hbar v_{\rm F}/(2a)$.

The computational results obtained using different theoretical approaches to the Casimir
force in graphene systems listed in Sec.~\ref{secKMM:2} were correlated with the results
found by means of the polarization tensor.\cite{KMM46} This was helpful in the determination
of the regions of applicability of each approach and in the justification of the results
using various phenomenological models. Specifically, it was shown that for a pristine
graphene the calculation approaches, which neglect the temperature dependence of its
dielectric response, could be applicable only at the shortest separations of about a few
angstr\"{o}ms. At the same time, the computational results using the polarization tensor
are in agreement\cite{KMM46} with nonrelativistic computations using
Coulomb coupling between density fluctuations with subsequent thermal averaging employed in the first publication which predicted the unusual
thermal effect in the Casimir force from graphene.\cite{KMM22}

Now we consider the impact of the nonzero gap and chemical potential on the size of thermal
correction to the Casimir pressure. Note that a nonzero energy gap and chemical potential
are inevitable with real graphene sheets particularly those on substrates as used in
experiments.\cite{KMM19,KMM23,KMM24}
As in experiments on
measuring the Casimir force from graphene one body is metallic, we consider an Au plate
interacting with the freestanding either pristine or real graphene sheet with
$\Delta=0.29~$eV and $\mu=0.24~$eV (these are the experimental parameters, see
Sec.~\ref{secKMM:5}). Computations of the thermal correction in each case are again
performed by (\ref{eqKMM1a}), (\ref{eqKMM2})--(\ref{eqKMM7}) and
(\ref{eqKMM11}). In both cases one should put $\Pi_{00,l}=\Pi_l=0$ in the reflection
coefficients $R_{\lambda}^{(1)}$ from (\ref{eqKMM2}) and substitute $\ve_l^{(1)}$ for Au
obtained from the optical data\cite{KMM47} extrapolated down to zero frequency\cite{KMM2,KMM3}
(the type of extrapolation does not influence the obtained results in this case due to
a smallness of the TE reflection coefficient for graphene at zero frequency).
As to the coefficients $R_{\lambda}^{(2)}$, one should put $\ve_l^{(2)}=1$ and use
$\Delta=\mu=0$ for a pristine graphene sheet and the specific values indicated above for
a real one.

\begin{figure}[b]
\vspace*{-9cm}
\centerline{\hspace*{-1.6cm}
\includegraphics[width=7.0in]{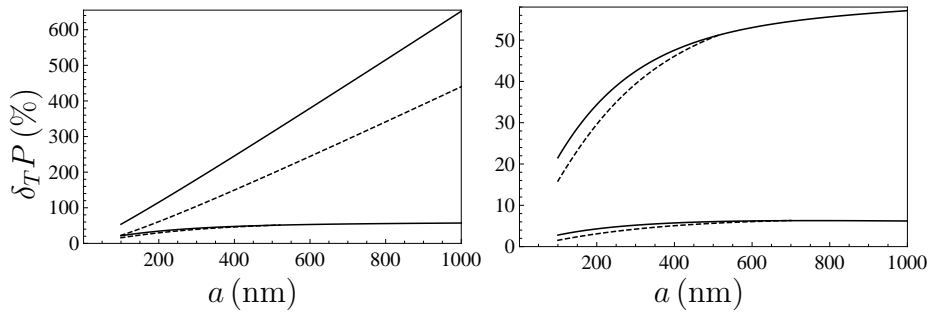}}
\vspace*{-12.cm}
\caption{
The relative thermal corrections to  the Casimir pressure between an Au plate and either
pristine or real freestanding graphene sheets are shown as functions of separation by the
top and bottom pairs of lines, respectively (left).
The relative thermal corrections to  the Casimir pressure between an Au plate and a
real graphene sheet deposited on a SiO${}_2$ plate are shown by the bottom pair of lines (right). In each pair, the solid line is computed at $T=300~$K using the full polarization tensor and the dashed line ---
its zero-temperature contribution taken at the Matsubara frequencies.
The top pair of lines (right) reproduces the bottom pair of lines (left).
\protect\label{figKMM:2}}
\end{figure}
The computational results for the relative thermal correction
to  the Casimir pressure between an Au plate and either pristine or real graphene sheet
are shown as the functions of separation in Fig.~\ref{figKMM:2} (left) by the
top and bottom pairs of lines, respectively. In each pair, the solid line indicates the
total thermal correction and the dashed line presents the implicit thermal correction with
neglected explicit dependence of polarization tensor on temperature as a parameter.
In this case the thermal effect originates entirely from a summation over the Matsubara
frequencies.

As is seen from the top solid line  in Fig.~\ref{figKMM:2} (left), for an Au plate
interacting with a pristine graphene sheet the thermal correction, though large enough,
is much smaller than for two pristine graphene sheets. At $a=100$, 200, 600, and 1000~nm,
it constitutes 53.7\%, 115.5\%, 379.5\%, and 659.9\% of the Casimir pressure calculated
at $T=0$, respectively. These should be compared with the respective values related to
the top line in Fig.~\ref{figKMM:1} (right) which are by almost a factor of 5 larger.
{}From the top dashed line in Fig.~\ref{figKMM:2} (left), one finds that at the same
respective separations the implicit thermal correction constitutes 22.5\%, 61.1\%, 244.2\%
and 439.9\%. Thus, in this case, the role of implicit correction increases with
increasing separation.

For an Au plate interacting with real graphene sheet the total thermal effect constitutes
21.5\%, 34.4\%, 53\%, and 58\% at $a=100$, 200, 600, and 1000~nm, respectively [see the
bottom solid line in Fig.~\ref{figKMM:2} (left)]. {}From the bottom dashed line one finds
15.9\%, 29.6\%,53\% and 58\% for the implicit thermal correction to the Casimir pressure
at $T=0$ at the same respective separations. This means that for a real freesranding
graphene sheet the role of explicit dependence of the polarization tensor on $T$ as a
parameter rapidly decreases with increasing separation.

Since in measurements of the Casimir force graphene is deposited on some substrate, we also
illustrate its impact on the size of the thermal correction to the Casimir pressure.
As a typical substrate, we consider a SiO${}_2$ plate with a sufficiently accurate
expression\cite{KMM48}  for the dielectric permittivity $\ve_l^{(2)}$. Now we perform
computations of the thermal correction to the Casimir pressure between an Au plate and
a real graphene sheet deposited on a SiO${}_2$ plate
by (\ref{eqKMM1a}), (\ref{eqKMM2})--(\ref{eqKMM7}) and
(\ref{eqKMM11}). In $R_{\lambda}^{(1)}$ from (\ref{eqKMM2}) we again
 put $\Pi_{00,l}=\Pi_l=0$  and use $\ve_l^{(1)}$ for Au.   In $R_{\lambda}^{(2)}$
we now use  $\ve_l^{(2)}$ for SiO${}_2$ and the experimental values of $\Delta$ and $\mu$
indicated above for a real graphene sheet.

The computational results are shown in Fig.~\ref{figKMM:2} (right) by the bottom solid line
as a function of separation. The relative thermal correction is equal to 2.79\%, 4.29\%,
6.2\%, and 6.2\% of the Casimir pressure at $T=0$ at $a=100$, 200, 600, and 1000~nm,
 respectively. The bottom dashed line presents similar results for the implicit thermal
 correction to the Casimir pressure at $T=0$. It is equal to 1.53\%, 3.10\%, 6.0\%, and
 6.2\% at the same respective separations. For comparison purposes,
in Fig.~\ref{figKMM:2} (right) we also reproduce the bottom pair of solid and dashed lines
from  Fig.~\ref{figKMM:2} (left) which shows the thermal corrections to the Casimir pressure
between an Au plate and a freestanding real graphene sheet. In Fig.~\ref{figKMM:2} (right)
this pair  takes the top position.

{}From Fig.~\ref{figKMM:2} one can conclude that nonzero values of the energy gap and
chemical potential of real graphene sample lead to a significant decrease of the thermal
effect at short separations as compared to the case of a pristine graphene sheet.
The thermal effect decreases further when the graphene sheet is deposited on a substrate
but still remains measurable in high precision experiments. Similar to the case of a
freestanding graphene sample, for a graphene-coated substrate the role of an explicit
contribution to the thermal effect rapidly decreases with increasing separation.

In the end of this section, we note that the thermal correction to the Casimir pressure
from graphene depends heavily \cite{KMM39} on whether the condition $\Delta<2\mu$ is
satisfied (as in our case) or $\Delta>2\mu$. Different aspects of the Casimir and
Casimir-Polder interactions from graphene are investigated in a number of papers using the
formalism of the polarization
tensor.
\cite{KMM49,KMM50,KMM51,KMM52,KMM53,KMM54,KMM55,KMM56,KMM57,KMM58,KMM59,KMM60,KMM60a,KMM61}

\section{First Experiment on Measuring the Casimir Interaction from Graphene}
\label{secKMM:4}

In this experiment,\cite{KMM19} the gradient of the Casimir force was measured between
an Au-coated hollow glass microsphere attached to an AFM cantilever and a graphene sheet
deposited on a SiO${}_2$ film covering a Si plate. The radius of the coated sphere was
$R=54.10\pm 0.09~\mu$m and the thickness of an Au coating was 280~nm allowing to consider
this sphere as all-gold. A large area graphene sample was obtained through a two-step
chemical vapor deposition process.\cite{KMM62} The grown graphene sheet was transferred
to a plate with $D=300~$nm thick SiO${}_2$ layer
on top of a Si substrate of $500~\mu$m thickness.

Measurements of the gradient of the Casimir force were performed at room temperature
$T=300~$K in high vacuum down to $10^{-9}~$Torr using the dynamic measurement scheme
employed earlier in measuring the Casimir interaction between metallic
surfaces.\cite{KMM63,KMM64} The total force
\begin{equation}
F_{\rm tot}(a,T)=F_{\rm el}(a)+F_{sp}(a,T)
\label{eqKMM12}
\end{equation}
\noindent
was the sum of an electric force $F_{\rm el}$ caused by the constant voltages $V_i$
applied to graphene sheet while the sphere remained grounded and the Casimir force $F_{sp}$
which depends on the temperature at the laboratory.

The force (\ref{eqKMM12}) leads to a modification of the resonant frequency of the
cantilever-sphere system from $\omega_0$ to some $\omega_r(a,T)$, and the frequency shift
\begin{equation}
\Delta\omega(a,T)=\omega_r(a,T)-\omega_0
\label{eqKMM13}
\end{equation}
\noindent
was measured by means of a phase-locked loop.

These measurements were performed at different separations in the linear regime of the
oscillator where the frequency shift is connected with the gradient of the total force
(\ref{eqKMM12}) according to \cite{KMM63,KMM65}
\begin{equation}
\Delta\omega(a,T)=-CF_{\rm tot}^{\prime}(a,T)=
-CF_{\rm el}^{\prime}(a)-CF_{sp}^{\prime}(a,T).
\label{eqKMM14}
\end{equation}

The calibration constant $C$ in (\ref{eqKMM14}) is given by $C=\omega_0/(2k)$ ($k$ is
the cantilever spring constant), $F_{\rm el}^{\prime}$ is a known function \cite{KMM63}
of the sphere radius, sphere-graphene separation $a=z_0+z_{\rm piezo}$, where
$z_{\rm piezo}$ is the distance moved by the graphene-coated SiO${}_2$-Si substrate, $z_0$
is the closest sphere-graphene separation, and of the residual potential difference $V_0$.
Note that the values of $z_0$, $V_0$, and $C$ are determined by means of electrostatic
calibration. Then, the experimental values of the gradient of the Casimir force between an
Au-coated sphere and a graphene-coated substrate are obtained from the measured frequency
shift using (\ref{eqKMM14})
\begin{equation}
F_{sp}^{\prime}(a,T)=-\frac{1}{C}\Delta\omega(a,T)-F_{\rm el}^{\prime}(a).
\label{eqKMM15}
\end{equation}

The comparison between experiment and theory was made by using the formalism presented in
Sec.~\ref{secKMM:2} and the proximity force approximation.
In this approximation,\cite{KMM3,KMM4} the Casimir force between a sphere and a plate is
expressed as
\begin{equation}
F_{sp}(a,T)=2\pi R{\cal F}(a,T),
\label{eqKMM16}
\end{equation}
\noindent
where the Casimir free energy per unit area of two parallel plates is given by
(\ref{eqKMM1}). By differentiating both sides of (\ref{eqKMM16}) with respect
to $a$, one expresses the gradient of the Casimir force between a sphere and a
graphene-coated substrate via the pressure between an Au plate and this substrate
\begin{equation}
F_{sp}^{\prime}(a,T)=-2\pi R{P}(a,T).
\label{eqKMM17}
\end{equation}

The error introduced from using the approximate equality (\ref{eqKMM17})
has been shown as
being less than $a/R$ based on the exact theory of the Casimir force using the gradient
expansion.\cite{KMM66,KMM67,KMM68} For the present experiment this error does not
exceed 0.5\% (see Sec.~\ref{secKMM:5} for a more detailed information).

The theoretical force gradients should also be corrected for the presence of surface
roughness.\cite{KMM3,KMM4,KMM69,KMM70,KMM70a}
In the framework of the multiplicative approach, which is sufficiently precise at short
separations used in this experiment,\cite{KMM3,KMM4} the gradient of the Casimir force
corrected for the presence of roughness is given by
\begin{equation}
F_{sp,{\rm theor}}^{\prime}(a,T)=\left(1+10
\frac{\delta_s^2+\delta_g^2}{a^2}\right)F_{sp}^{\prime}(a,T).
\label{eqKMM18}
\end{equation}
\noindent
Here, the root-mean-square roughness amplitudes on the surfaces of the sphere and graphene
measured by means of the usual AFM with a sharp tip are equal to $\delta_s=1.6\pm0.1~$nm
and $\delta_g=1.5\pm0.1~$nm, respectively. Thus, the maximum contribution of the surface
roughness is equal to only 0.1\% at the shortest separation in this experiment.

Computations of the Casimir pressure in (\ref{eqKMM17}) were performed by (\ref{eqKMM1a})
at $T=300~$K as follows.\cite{KMM20} The reflection coefficients $R_{\lambda}^{(1)}$ on an Au surface are given by (\ref{eqKMM2}) where one should put $\Pi_{00,l}=\Pi_l=0$ and use
$\ve_l^{(1)}$ for Au (see Sec.~\ref{secKMM:3}).
The reflection coefficients $R_{\lambda}^{(2)}$ in (\ref{eqKMM1})  should be replaced with
\begin{equation}
\widetilde{R}_{\lambda}^{(2)}\xkb=
\frac{R_{\lambda}^{(2)}\xkb+r_{\lambda}^{(2)}\xkb\, e^{-2Dk_l^{(2)}}}{1+
R_{\lambda}^{(2)}\xkb r_{\lambda}^{(2)}\xkb\, e^{-2Dk_l^{(2)}}}
\label{eqKMM19}
\end{equation}
\noindent
because in this experiment graphene is deposited on a SiO${}_2$ film of thickness $D$
covering the Si plate.

Here, the coefficients $R_{\lambda}^{(2)}$ are defined in (\ref{eqKMM2}) where the
permittivity
$\ve_l^{(2)}$ for SiO${}_2$ was discussed in Sec.~\ref{secKMM:3}.
The polarization tensor $\Pi_{00,l}$ and $\Pi_l$ entering $R_{\lambda}^{(2)}$ is given in
(\ref{eqKMM4})--(\ref{eqKMM7}). It was used\cite{KMM20} with $\mu=0$ and $\Delta$ varying
from 0 to 0.1~eV (the exact values for the experimental graphene sample were not determined).
As to the reflection coefficients  $r_{\lambda}^{(2)}$ in (\ref{eqKMM19}), they describe
the reflection on the boundary plane between semispaces made of SiO${}_2$ and Si and have
the standard form
\begin{equation}
r_{\rm TM}^{(2)}\xkb=
\frac{\ve_l^{\rm Si}k_l^{(2)}-\ve_l^{(2)}k_l^{\rm Si}}{\ve_l^{\rm Si}k_l^{(2)}+
\ve_l^{(2)}k_l^{\rm Si}},
\qquad
r_{\rm TE}^{(2)}\xkb=
\frac{k_l^{(2)}-\ve_l^{(2)}k_l^{\rm Si}}{k_l^{(2)}+\ve_l^{(2)}k_l^{\rm Si}},
\label{eqKMM20}
\end{equation}
\noindent
where $k_l^{\rm Si}=(\skb+\ve_l^{\rm Si}\xi_l^2/c^2)^{1/2}$.

The dielectric permittivity of Si along the imaginary frequency axis,
$\ve_l^{\rm Si}=\ve^{\rm Si}(i\xi_l)$, added some uncertainty in the theoretical analysis
of this experiment. The B-doped Si plate used\cite{KMM19} had a nominal resistivity
between 0.001 and 0.005~$\Omega$\,cm. This leads\cite{KMM71} to a density of charges
varying from $1.6\times 10^{19}$ to $7.8\times 10^{19}~\mbox{cm}^{-3}$ which is above the
critical value $3.95\times 10^{18}~\mbox{cm}^{-3}$ at which the dielectric-to-metal
transition occurs.\cite{KMM72} The resulting plasma frequency $\omega_p$ varied in the
relatively wide range\cite{KMM19} between $5\times 10^{14}$ and $11\times 10^{14}~$rad/s
and was used in the extrapolation of the Si optical data\cite{KMM47} to zero frequency.

\begin{figure}[b]
\vspace*{-9cm}
\centerline{\hspace*{-1.5cm}
\includegraphics[width=7in]{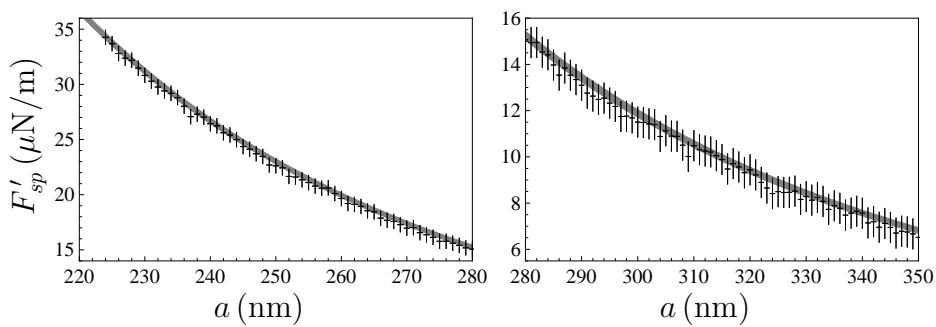}}
\vspace*{-12.cm}
\caption{The gradients of the Casimir force between an Au-coated sphere and a
graphene sheet deposited on a SiO${}_2$ film covering a Si plate computed at
$T=300~$K using the polarization tensor are shown as  functions of separation by the
dark gray band over the intervals from 224 to 280~nm (left) and
from 280 to 350~nm (right). The measurement data are indicated as  crosses.
\protect\label{figKMM:3}}
\end{figure}
In Fig.~\ref{figKMM:3}, the measured gradients of the Casimir force from (\ref{eqKMM15})
between an Au-coated sphere and graphene-coated substrate are shown as crosses. The arms
of the crosses indicate the total experimental errors in measuring the force gradients
and separations determined at the 67\% confidence level. The theoretical gradients
of the Casimir force computed as a function of separation by (\ref{eqKMM17}) and
(\ref{eqKMM18}), as explained above, are shown as the dark gray bands over the intervals
from 224 to 280~nm (left) and from 280 to 350~nm (right). The width of the bands is
mostly determined by the uncertainties in the values of $\omega_p$ for the Si plate and
in the energy gap $\Delta$ of graphene.

Over the entire measurement range from 224 to 500~nm (Fig.~\ref{figKMM:3} illustrates
only a part of this interval) the theoretical force gradients were found to be in good
agreement with the predictions of the Lifshitz theory describing the electromagnetic
response of graphene by means of the polarization tensor.\cite{KMM20}
Note that theoretical predictions of the same theory using the reflection coefficients
found in the framework of the hydrodynamic model of graphene are excluded by the measurement
data of this experiment.\cite{KMM73} The question remains of what this experiment says about
the large thermal effect in the Casimir force from graphene which, according to
Sec.~\ref{secKMM:3}, should exist and be observable at short separations below $1~\mu$m.

\begin{figure}[b]
\vspace*{-8.5cm}
\centerline{\hspace*{1.5cm}
\includegraphics[width=7in]{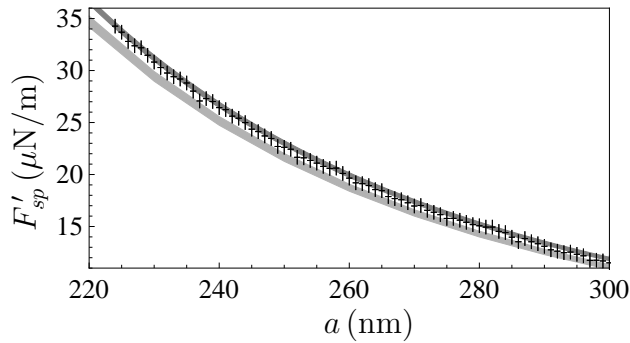}}
\vspace*{-12.cm}
\caption{The gradients of the Casimir force between an Au-coted sphere and a graphene
sheet deposited on a SiO${}_2$ film covering a Si plate computed at $T=300~$K and $T=0~$K
using the polarization tensor are shown as  functions of separation by the top
(dark gray) and bottom (light gray) bands, respectively.
The measurement data are indicated as  crosses.
\protect\label{figKMM:4}}
\end{figure}
The question above was investigated\cite{KMM21} taking into account
the experimental uncertainties
and the material properties of the substrate supporting the graphene sheet. The gradient of the
Casimir force at zero temperature was computed using the same formalism as above but
replacing the summation in (\ref{eqKMM1a}) with integration along the imaginary frequency
axis in accordance with (\ref{eqKMM10}). The obtained computational results are shown in
Fig.~\ref{figKMM:4} by the bottom (light gray) band as a function of separation.
The width of this band was found using the same uncertainties as considered above in
computations of the Casimir force at $T=300~$K. The theoretical band computed at $T=300~$K
is reproduced from Fig.~\ref{figKMM:3} as the top (dark gray) band together with the
measurement data indicated as the crosses. As can be seen in Fig.~\ref{figKMM:4}, the
thickness of the bottom band computed at $T=0~$K is somewhat larger than of the top one
computed at $T=300~$K. This is because under the condition $\Delta>2\mu$ (later\cite{KMM39}
the value of the chemical potential in this experiment was estimated as
$\mu\approx0.02~$eV) an impact of the nonzero energy gap $\Delta$,
which can be as much as 0.1~eV, on the force gradient is stronger at zero temperature.

{}From Fig.~\ref{figKMM:4} it is seen that the bottom theoretical band computed at $T=0~$K
is slightly below the measurement data which, however, touch it in a number of data points.
This suggests that there is some evidence in favor of the presence of thermal effect which,
nevertheless, cannot be considered as a solid confirmation for its existence.

According to the results obtained,\cite{KMM21}  an impact of graphene deposited on a
dielectric substrate on the thermal Casimir force increases with decreasing dielectric
permittivity of the substrate material. This means that the SiO${}_2$ substrate is
appropriate for observation of the thermal effect in the Casimir force from graphene.
It was concluded, however, that to attain this goal it is necessary to increase the
thickness of the SiO${}_2$ film up to at least $2~\mu$m in order to eliminate the
detrimental impact of the Si substrate. Following this analysis,
it has become possible to demonstrate the large thermal effect
in the next experiment on measuring the Casimir interaction
from graphene.

\section{Demonstration of Large Thermal Effect in the Casimir Interaction from Graphene}
\label{secKMM:5}

In the second experiment on measuring the Casimir interaction
from graphene,\cite{KMM23,KMM24} an Au-coated hollow glass microsphere of radius
$R=60.35\pm 0.5~\mu$m was used as the first test body, whereas the second one was a graphene
sheet deposited on top of thick SiO${}_2$ substrate. Thus, the Si plate, which added
uncertainty to the comparison between experiment and theory in the first experiment
(see Sec.~\ref{secKMM:4}), was removed.

The graphene sheet used was made from a large-area graphene sample grown on a Cu foil
using the method of chemical vapor deposition.\cite{KMM75}
This sheet was transferred onto an optically polished SiO${}_2$ substrate of 10~cm diameter
and 0.05~cm thickness \cite{KMM76} using an electrochemical delamination
procedure.\cite{KMM75,KMM77}

An important new feature of the second experiment, as compared to the first one, is that
a small but nonzero mass of electronic quasiparticles in graphene, which leads to an
energy gap $\Delta$ in their spectrum,\cite{KMM12,KMM13} was directly measured rather than
estimated in some qualitative manner. This was made by means of scanning tunneling
spectroscopy \cite{KMM78} and resulted in the value $\Delta= 0.29\pm 0.05~$eV exceeding
the one estimated in the first experiment.

Another important feature addressed in the second experiment is that any real graphene
sheet contains some fraction of impurities and, as a result, is characterized by a nonzero
value of the chemical potential.\cite{KMM12,KMM13} The polarization tensor of graphene,
taking the chemical potential of graphene $\mu$ into account, was derived\cite{KMM18}
only in 2016. Because of this, in the course of first experiment it was not possible
to reliably calculate the impact of impurities on the gradient of the Casimir force.

In the second experiment, the mean concentration of impurities in the graphene sheet was
measured by Raman spectroscopy\cite{KMM79} with the result
$\bar{n}=(4.2\pm 0.3)\times10^{12}~\mbox{cm}^{-2}$. Due to the transfer process used the
expected dominant type of impurities was Na. Then the value of the chemical potential
at $T=0~$K was found\cite{KMM80}
\begin{equation}
\mu=\hbar v_{\rm F}\sqrt{\pi\bar{n}}=0.24\pm0.01~\mbox{eV}.
\label{eqKMM21}
\end{equation}
\noindent
This value can also be used at room temperature because the relatively large chemical
potential $\mu=0.24~$eV is almost temperature-independent.\cite{KMM81}

The measurement scheme of the gradient of the Casimir force was the same as in the first
experiment described in Sec.~\ref{secKMM:4}. This means that measurements were performed
in high vacuum at $T=294.0\pm 0.5~$K temperature by using a modified AFM
cantilever-based technique operated in the dynamic mode. There was, however, an important
improvement allowing a significant decrease of the total experimental error in measuring
the force gradient. The AFM cantilever spring constant $k$ was reduced through chemical
etching (like it was done in a recent experiment with metallic test bodies\cite{KMM9})
leading to the corresponding decrease of the resonant frequency of the cantilever $\omega_0$.
As a result, the value of the calibration constant $C$ in (\ref{eqKMM15}) was increased by
up to a factor 8 leading to a significant decrease of the total experimental error from
$0.64~\mu$N/m in the first experiment (see Fig.~\ref{figKMM:3}) to $0.14~\mu$N/m in the
second.

Computations of the gradients of the Casimir force were performed by (\ref{eqKMM1a})
using the reflection coefficients (\ref{eqKMM2}), the polarization tensor
(\ref{eqKMM4})--(\ref{eqKMM7}), and the equalities (\ref{eqKMM17}) and (\ref{eqKMM18}).
In the reflection coefficients $R_{\lambda}^{(1)}$ one should put $\Pi_{00,l}=\Pi_l=0$
and take for $\ve_l^{(1)}$ the values of the dielectric permittivity of Au at pure
imaginary Matsubara frequencies. In $R_{\lambda}^{(2)}$ the polarization tensor at
$T=294~$K is taken with $\Delta=0.29~$eV and $\mu=0.24~$eV as was measured for a graphene
sheet used and the dielectric permittivity $\ve_l^{(2)}$ refers to a SiO${}_2$ substrate.
The root-mean-square roughness amplitudes on the surfaces of a sphere and graphene were
measured to be $\delta_s=0.9\pm0.1~$nm and $\delta_g=1.5\pm0.1~$nm, respectively.

\begin{figure}[b]
\vspace*{-9cm}
\centerline{\hspace*{-1.4cm}
\includegraphics[width=7in]{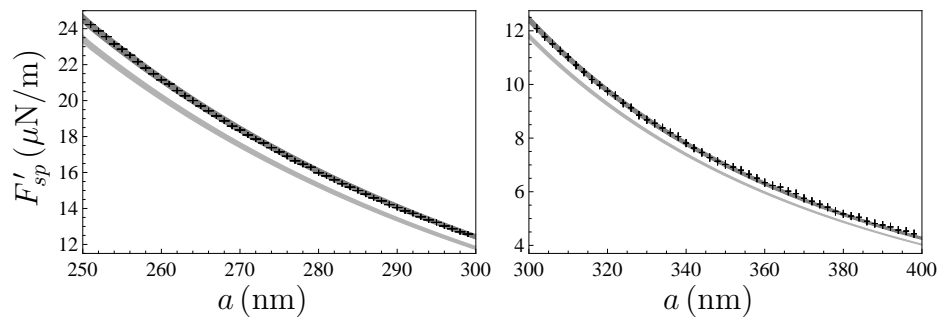}}
\vspace*{-12.cm}
\caption{The gradients of the Casimir force between an Au-coted sphere and a
graphene sheet deposited on a SiO${}_2$ substrate computed at $T=294~$K and $T=0~$K
are shown as  functions of separation by the top (dark gray) and bottom
 (light gray) bands, respectively, over the intervals from 250 to 300~nm (left) and
from 300 to 400~nm (right). The measurement data are indicated as  crosses.
\protect\label{figKMM:5}}
\end{figure}
The computational results for the gradient of the Casimir force are shown in
Fig.~\ref{figKMM:5} as the functions of separation by the top (dark gray) bands
over the intervals from 250 to 300~nm (left) and from 300 to 400~nm (right).
The width of these bands is determined in the following conservative way.
The upper boundary lines of the theoretical bands were computed with the largest
allowed value of $\mu=0.25~$eV and the smallest allowed value of $\Delta=0.24~$eV.
The lower boundary lines were computed with the smallest
allowed value of $\mu=0.23~$eV and the largest allowed value of $\Delta=0.34~$eV.
This is because an increase of $\mu$ with fixed $\Delta$ increases the force gradient,
whereas an increase of $\Delta$ with $\mu={\rm const}$ decreases the force
gradient.\cite{KMM39}
The width of the theoretical band was also increased to take into account the 0.5\%
errors arising from uncertainties in the optical data of Au and SiO${}_2$ and the error
in the sphere radius indicated above.

As was mentioned in Sec.~\ref{secKMM:4}, the theoretical force gradients computed
by (\ref{eqKMM17}) are burdened with an error due to the deviations from PFA.
According to the results obtained using the gradient expansion\cite{KMM66,KMM67,KMM68,KMM82}
and the scattering approach,\cite{KMM83,KMM84,KMM85,KMM86}  in
the sphere-plate geometry the PFA leads to slightly
larger force gradients than are given by the exact computations using these methods. Because of this, the upper boundary lines of the top theoretical
bands in Fig.~\ref{figKMM:5} remained as they are with no correction for the PFA error.
As to the lower lines bounding the theoretical bands, they are corrected for a maximum
possible correction factor of $(1-a/R)$.

The experimental gradients of the Casimir force are shown in Fig.~\ref{figKMM:5} as the
crosses whose arms indicate the total experimental errors determined at the 67\% confidence
level. It is seen that the measurement data are in good agreement with theoretical
predictions of the top band computed using the polarization tensor at $T=294~$K with
the measured values of $\Delta$ and $\mu$. Similar good agreement holds\cite{KMM23,KMM24}
in the remaining range of experimental separations from 400 to 700~nm which is not shown
in Fig.~\ref{figKMM:5}.

The computations of the gradient of the Casimir force were repeated at $T=0~$K.
In so doing, a summation in the Matsubara frequencies in (\ref{eqKMM1a}) was replaced with
an integration over continuous $\xi$ in accordance to (\ref{eqKMM10}).
It is important to stress that for a real graphene sheet with $\Delta<2\mu$ (as in this
experiment) the polarization tensor at $T=0~$K is not given by (\ref{eqKMM5}) but is
defined by the equalities
\begin{eqnarray}
&&
\Pi_{00}(i\xi,\kb,0,\Delta,\mu)=\Pi_{00}^{(0)}(i\xi,\kb,\Delta)+
\lim_{T\to 0}\Pi_{00}^{(1)}(i\xi,\kb,T,\Delta,\mu),
\nonumber \\
&&
\Pi(i\xi,\kb,0,\Delta,\mu)=\Pi^{(0)}(i\xi,\kb,\Delta)+
\lim_{T\to 0}\Pi^{(1)}(i\xi,\kb,T,\Delta,\mu).
\label{eqKMM22}
\end{eqnarray}
\noindent
As a result, $\Pi_{00}$ and $\Pi$ at $T=0~$K depend both on $\Delta$ and $\mu$ in this case
[under the condition $\Delta>2\mu$ the polarization tensor at $T=0~$K is given\cite{KMM39}
by (\ref{eqKMM5}) and, thus, does not depend on $\mu$].

The computational results for the gradient of the Casimir force at $T=0~$K are shown
in Fig.~\ref{figKMM:5} as the functions of separation by the bottom (light gray) bands
over the intervals from 250 to 300~nm (left) and from 300 to 400~nm (right).
The thickness of these bands is computed in the same conservative way as  the top
ones taking into account all theoretical uncertainties and errors as explained above.
According to Fig.~\ref{figKMM:5}, the top bands do not intersect with the bottom ones
demonstrating the predicted large thermal effect from graphene at short separations.
The same holds\cite{KMM23,KMM24} at separations up to $a=517~$nm (at larger separations
both bands overlap). Thus, this experiment demonstrates the presence of the thermal effect
in the gradient of the Casimir force from graphene sample in the region of relatively
short separations from 250 to 517~nm.

\begin{figure}[h]
\vspace*{-8.5cm}
\centerline{\hspace*{1.5cm}
\includegraphics[width=7in]{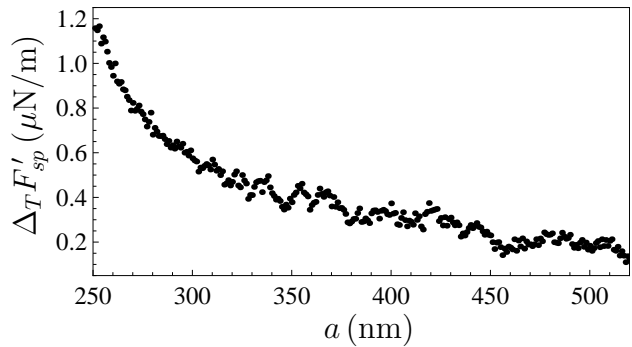}}
\vspace*{-12.cm}
\caption{The observed thermal correction to the gradient of the Casimir force between
an Au-coated sphere and a graphene-coated SiO${}_2$ substrate
is shown as a function of separation.
\protect\label{figKMM:6}}
\end{figure}

To explicitly demonstrate the observed thermal effect, in Fig.~\ref{figKMM:6} we show
the thermal correction to the gradient of the Casimir force equal to a difference between
the mean measured gradients presented in Fig.~\ref{figKMM:5} and the theoretical ones
computed at $T=0~$K as a function of separation. This correction constitutes 4\%, 5\%,
7\%, and 8.5\% of the total gradient of the Casimir force at separations $a=250$, 300,
400, and 500~nm, respectively. As is seen in Fig.~\ref{figKMM:6}, its absolute values
are significantly larger than the total error in measuring the gradient of the Casimir
force equal to $0.14~\mu$N/m.

\section{Conclusions and Discussion}
\label{secKMM:6}

In the foregoing, we have considered the Casimir effect from graphene. This is a novel
material which has already demonstrated several unusual features of its mechanical,
electrical, and optical properties. As mentioned in Sec.~\ref{secKMM:1}, the striking
feature of the Casimir effect between two pristine graphene sheets is a predicted\cite{KMM22}
huge thermal correction equal to thousands of percent of the zero-temperature pressure even
at short separations of a few hundred nanometers. This effect should be compared with a few
percent thermal correction to the Casimir pressure between two parallel metallic plates
described by the Drude model which was excluded by the results of many high precision
experiments.

As discussed in Sec.~\ref{secKMM:2}, the Casimir effect from graphene at separations exceeding
100~nm can be reliably described by using the Lifshitz theory where the dielectric response
of graphene is expressed via the polarization tensor at nonzero temperature
accounting for the nonzero energy gap and chemical potential. Taking into consideration that
the explicit expression for the polarization tensor is derived from the first principles
of quantum electrodynamics at nonzero temperature and provides the correct description of the
electrical and optical properties of graphene,\cite{KMM87,KMM88} the crucial question
arises whether the predicted huge thermal effect does occur in nature.

In Sec.~\ref{secKMM:3}, it is shown that for real graphene sheets possessing nonzero energy
gap and chemical potential the predicted effect is significantly suppressed. In the case
of graphene-coated dielectric substrate interacting with a metallic plate at separations of
a few hundred nanometers, its size is reduced to a few percent, i.e., to the same value as
an experimentally excluded thermal correction for metals described by the Drude model.
At the same time, the Lifshitz theory using the dielectric response of graphene given
by the polarization tensor was found to be in agreement with the Nernst heat theorem
which is not the case for an idealized model of Drude metals with perfect crystal lattices
and for dielectric materials described by the frequency-dependent dielectric permittivity
with inclusion of the conductivity at nonzero temperature. This adds 
considerable significance to the
experimental investigation of the Casimir effect from graphene.

The first experiment on measuring the Casimir interaction from graphene,\cite{KMM19}
considered  in Sec.~\ref{secKMM:4}, was found\cite{KMM20} in good agreement with theoretical
predictions of the Lifshitz theory using the polarization tensor. Taking into account the
relatively large experimental and theoretical errors from the not fully characterized graphene
sample and the two-layer substrate, and the unavailability
of the polarization tensor accounting
for the chemical potential at that time, it was not possible to reliably separate
only a few percent thermal effect from the total measured force gradient.

These problems were successfully solved in the second, refined, experiment on measuring
the gradient of the Casimir force between an Au-coated sphere and a graphene-coated
SiO${}_2$ substrate discussed in Sec.~\ref{secKMM:5}. In this experiment,\cite{KMM23,KMM24}
the single-layer thick dielectric substrate has been used made of a material with low static
dielectric permittivity which is favorable for an observation of the thermal
effect.\cite{KMM21} What is more, the graphene sheet was carefully characterized by
performing independent measurements of its energy gap and chemical potential.
By reducing the spring constant of the AFM cantilever through chemical etching, the
calibration constant of the oscillator was increased by up to a factor of 8.
In the end, the total experimental error in the second experiment was reduced by the factor
of 4.6 as compared to the first one.

Finally, the theoretical gradients of the Casimir force in the experimental configuration
have been computed using the Lifshitz theory and the exact polarization tensor of graphene
taking into account the nonzero values of its energy gap and chemical potential.
The theoretical values at the experimental temperature $T=294~$K were found in good agreement
with the measurement data over the entire range of separations from 250 to 700~nm with no
fitting parameters.  The comparison of the same data with the theoretical force gradients
computed at $T=0~$K unambiguously demonstrated the presence of the predicted thermal effect
which size ranges from 4\% to 8.5\% of the total force gradient when separation increases
from 250 to 500~nm, respectively.

Thus, the Lifshitz theory using the exact response functions of graphene to electromagnetic
fluctuations does not have any problems being in good agreement with both the measurement
data and with the principles of thermodynamics. The long-term problems in theoretical
description of the Casimir force between metallic surfaces mentioned  in Sec.~\ref{secKMM:1}
arise when a metal is described by the Drude model which has been 
carefully tested in the area of
propagating waves. Because of this, one may suggest that this phenomenological model fails
to provide an adequate description of the electromagnetic response to evanescent waves
which are off the mass shell in the free space but
make an important contribution to the
Casimir force. Taking into account that the polarization
tensor of graphene is equivalent to the spatially nonlocal dielectric permittivities,
the Drude-like phenomenological nonlocal response functions were
proposed\cite{KMM89,KMM90,KMM91,KMM92} which nearly coincide with the Drude model in the
area of propagating waves but bring the Lifshitz theory in agreement with both thermodynamics
and experiment for the test bodies made of nonmagnetic and magnetic metals.

Hence, the fundamental description of an electromagnetic response in the case of 3D material
bodies (an attempt in this direction was undertaken recently\cite{KMM93}), as has been
 made for graphene, may be also helpful in solving the complicated problems
of  Casimir physics for usual materials.

\section*{Acknowledgments}

The work of U.~M.~was partially supported by the NSF grant PHY-2012201.
The work of G.~L.~K. and V.~M.~M.  was partially
supported by the Peter the Great Saint Petersburg Polytechnic
University in the framework of the Russian state assignment for basic research
(project No.\ FSEG-2020-0024).
This paper has been supported by the  Kazan Federal University
Strategic Academic Leadership Program.


\begin{thebibliography}{999}
\bibitem{KMM1}
H.~B.~G.~Casimir,
{\it Proc. Kon. Ned. Akad. Wet. B} {\bf 51}, 793 (1948).
\bibitem{KMM2}
E.~M.~Lifshitz,
{\it Zh. Eksp. Teor. Fiz.} {\bf 29}, 94 (1955)
[{\it Sov. Phys. JETP} {\bf 2}, 73 (1956)].
\bibitem{KMM2a}
E.~M.~Lifshitz and L.~P.~Pitaevskii,
{\it Statistical Physics, Part II}
(Pergamon, Oxford, 1980).
\bibitem{KMM3}
G.~L.~Klimchitskaya, U. Mohideen and V.\ M.\ Mostepanenko,
{\it Rev. Mod. Phys.} {\bf 81}, 1827 (2009).
\bibitem{KMM4}
M.~Bordag, G.~L.~Klimchitskaya, U.\ Mohideen and
V.\ M.\ Mostepanenko, {\it Advances in the Casimir Effect}
(Oxford University Press, Oxford, 2015).
\bibitem{KMM5}
L.~M.~Woods, D.~A.~R.~Dalvit, A.~Tkatchenko, P.\ Rodriguez-Lopez,
A.\ W.\ Rodriguez and R.\ Podgornik,
{\it Rev. Mod. Phys.} {\bf 88}, 045003 (2016).
\bibitem{KMM6}
V.~M.\ Mostepanenko,
{\it Universe} {\bf 7}, 84 (2021).
\bibitem{KMM7}
G.~Bimonte, D.~L\'{o}pez and R.~S.\ Decca,
{\it Phys. Rev. B} {\bf 93}, 184434 (2016).
\bibitem{KMM8}
M.~Liu, J.~Xu,
G.~L.~Klimchitskaya, V.~M.\ Mostepanenko and U.\ Mohideen,
{\it Phys. Rev. B} {\bf 100}, 081406(R) (2019).
\bibitem{KMM9}
M.~Liu, J.~Xu,
G.~L.~Klimchitskaya, V.~M.\ Mostepanenko and U.\ Mohideen,
{\it Phys. Rev. A} {\bf 100}, 052511 (2019).
\bibitem{KMM10}
G. Bimonte, B. Spreng, P.~A.~Maia Neto, G.-L.~Ingold, G.~L.~Klimchitskaya,
V.~M.\ Mostepanenko and R.~S.~Decca,
{\it Universe} {\bf 7}, 93 (2021).
\bibitem{KMM11}
A.~H.~Castro Neto, F.~Guinea, N.~M.~R.~Peres, K.~S.~Novoselov
and A.~K.~Geim,
{\it Rev. Mod. Phys.} {\bf 81}, 109 (2009).
\bibitem{KMM11a}
N.~M.~R.~Peres,
{\it Rev. Mod. Phys.} {\bf 82}, 2673 (2010).
\bibitem{KMM11b}
S.~Das~Sarma, S.~Adam, E.~H.~Hwang and E.\ Rossi,
{\it Rev. Mod. Phys.} {\bf 83}, 407 (2011).
\bibitem{KMM12}
 H.\ Aoki and M.\ S.\ Dresselhaus (eds.),
{\it Physics of Graphene}
(Springer, Cham, 2014).
\bibitem{KMM13}
M.~I.~Katsnelson,
{\it The Physics of Graphene}
(Cambridge University Press, Cambridge, 2020).
\bibitem{KMM14}
T.~Zhu, M.~Antezza and J.-S.~Wang,
{\it Phys. Rev. B} {\bf 103}, 125421 (2021).
\bibitem{KMM15}
M.~Bordag, I.~V.~Fialkovsky, D.~M.~Gitman and
D.~V.~Vassilevich,
{\it Phys. Rev. B} {\bf 80}, 245406 (2009).
\bibitem{KMM16}
I.~V.~Fialkovsky, V.~N.~Marachevsky and D.~V.~Vassilevich,
{\it Phys. Rev. B} {\bf 84}, 035446 (2011).
\bibitem{KMM17}
M.~Bordag, G.~L.~Klimchitskaya, V.~M.~Mostepanenko and V.~M.~Petrov,
{\it Phys. Rev. D} {\bf 91}, 045037 (2015)
[Erratum: {\it ibid.} {\bf 93}, 089907 (2016)].
\bibitem{KMM18}
M.~Bordag, I.~Fialkovskiy and D.~Vassilevich,
{\it Phys. Rev. B} {\bf 93}, 075414 (2016)
[Erratum: {\it ibid.} {\bf 95}, 119905 (2017)].
\bibitem{KMM18a}
M.~Dressel and G.~Gr\"{u}ner,
{\it Electrodynamics of Solids: Optical Properties of Electrons in Metals}
(Cambridge University Press, Cambridge, 2003).
\bibitem{KMM19}
A.~A.~Banishev, H.~Wen, J.~Xu, R.~K.~Kawakami, G.~L.~Klimchitskaya,
V.~M.~Mostepanenko and U.~Mohideen,
{\it Phys. Rev. B} {\bf 87}, 205433 (2013).
\bibitem{KMM20}
G.~L.~Klimchitskaya, U.~Mohideen and V.~M.~Mostepanenko,
{\it Phys. Rev. B} {\bf 89}, 115419 (2014).
\bibitem{KMM21}
G.~L.~Klimchitskaya and V.~M.~Mostepanenko,
{\it Phys. Rev. A} {\bf 89}, 052512 (2014).
\bibitem{KMM22}
G.~G\'{o}mez-Santos,
{\it Phys. Rev. B} {\bf 80}, 245424 (2009).
\bibitem{KMM23}
M. Liu, Y. Zhang, G. L. Klimchitskaya, V. M. Mostepanenko and  U. Mohideen,
{\it Phys. Rev. Lett.} {\bf 126}, 206802 (2021).
\bibitem{KMM24}
M. Liu, Y. Zhang, G. L. Klimchitskaya, V. M. Mostepanenko and  U. Mohideen,
{\it Phys. Rev. B} {\bf 104}, 085436 (2021).
\bibitem{KMM24a}
O.~Kenneth and I.~Klich,
{\it Phys. Rev. B} {\bf 78}, 014103 (2008).
\bibitem{KMM24b}
T.~Emig, N.~Graham, R.~L.~Jaffe and M.~Kardar,
{\it Phys. Rev. D} {\bf 77}, 025005 (2008).
\bibitem{KMM25}
S. J.~Rahi, T.~Emig, N.~Graham, R. L.~Jaffe and M.~Kardar,
{\it Phys. Rev. D} {\bf 80}, 085021 (2009).
\bibitem{KMM26}
G.~Barton,
{\it J. Phys. A: Math. Gen.} {\bf 38}, 2997 (2005).
\bibitem{KMM27}
M.~Bordag,
{\it J. Phys. A: Math.~Gen.} {\bf 39}, 6173 (2006).
\bibitem{KMM28}
M.~Bordag, B.~Geyer, G.~L.~Klimchitskaya and V.~M.~Mostepanenko,
{\it Phys. Rev. B} {\bf 74}, 205431 (2006).
\bibitem{KMM29}
L.~A.~Falkovsky and A.~A.~Varlamov,
{\it Eur. Phys. J. B} {\bf 56}, 281 (2007).
\bibitem{KMM30}
L.~A.~Falkovsky and S.~S.~Pershoguba,
{\it Phys. Rev. B} {\bf 76}, 153410 (2007).
\bibitem {KMM31}
D.~Drosdoff, A.~D.~Phan, L.~M.~Woods, I.\ V.\ Bondarev
and J.\ F.\ Dobson,
{\it Eur. Phys. J. B} {\bf 85}, 365 (2012).
\bibitem{KMM32}
J.~F.~Dobson, A.~White and A.~Rubio,
{\it Phys. Rev. Lett.} {\bf 96}, 073201 (2006).
\bibitem{KMM33}
T.~Stauber, N.~M.~R.~Peres and A.~K.~Geim,
{\it Phys. Rev. B} {\bf 78}, 085432 (2008).
\bibitem{KMM34}
Bo~E.~Sernelius,
{\it Europhys. Lett.} {\bf 95}, 57003 (2011).
\bibitem{KMM35}
J.~Sarabadani, A.~Naji, R.~Asgari and R.~Podgornik,
{\it Phys. Rev. B} {\bf 84}, 155407 (2011);
{\it Phys. Rev. B} {\bf 87}, 239905(E) (2013).
\bibitem{KMM36}
Bo E. Sernelius,
{\it Phys. Rev. B} {\bf 85}, 195427 (2012); {\bf 89}, 079901(E) (2014).
\bibitem{KMM36a}
C.~Abbas, B.~Guizal and M.~Antezza,
{\it Phys. Rev. Lett.} {\bf 118}, 126101 (2017).
\bibitem{KMM37}
D.~Drosdoff and L.~M.~Woods,
{\it Phys. Rev. B} {\bf 82}, 155459 (2010).
\bibitem{KMM37a}
D.~Drosdoff and L.~M.~Woods,
{\it Phys. Rev. A} {\bf 84}, 062501 (2011).
\bibitem{KMM37b}
A.~D.~Phan, N.\ A.\ Viet, N.\ A.\ Poklonski, L.~M.~Woods
and  C.\ H.\ Le,
{\it Phys. Rev. B} {\bf 86}, 155419 (2012).
\bibitem{KMM38}
G.~L.~Klimchitskaya and V.~M.~Mostepanenko,
{\it Phys. Rev. B} {\bf 91}, 174501 (2015).
\bibitem{KMM39}
G.~Bimonte, G.~L.~Klimchitskaya and V.~M.~Mostepanenko,
{\it Phys. Rev. B} {\bf 96}, 115430 (2017).
\bibitem{KMM40}
G.~L.~Klimchitskaya, V.~M.~Mostepanenko and
Bo~E.~Sernelius,
{\it Phys. Rev. B} {\bf 89}, 125407 (2014).
\bibitem{KMM41}
V.~B.~Bezerra, G.~L.~Klimchitskaya,
V.~M.~Mostepanenko and C.~Romero,
{\it Phys. Rev. A} {\bf 94}, 042501 (2016).
\bibitem{KMM42}
G.~L.~Klimchitskaya and V.~M.~Mostepanenko,
{\it Phys. Rev. D} {\bf 102}, 016006 (2020).
\bibitem{KMM43}
G.~L.~Klimchitskaya and V.~M.~Mostepanenko,
{\it Phys. Rev. A} {\bf 98}, 032506 (2018).
\bibitem{KMM44}
G.~L.~Klimchitskaya and V.~M.~Mostepanenko,
{\it Phys. Rev. D} {\bf 101}, 116003 (2020).
\bibitem{KMM45}
G.~L.~Klimchitskaya and V.~M.~Mostepanenko,
{\it Universe} {\bf 6}, 150 (2020).
\bibitem{KMM46}
G.~L.~Klimchitskaya
and V.~M.~Mostepanenko,
{\it Phys. Rev. B} {\bf 87}, 075439 (2013).
\bibitem{KMM47}
E.~D.~Palik (ed.),
{\it Handbook of Optical Constants of Solids}
(Academic Press, New York, 1985).
\bibitem{KMM48}
L. Bergstr\"{o}m,
{\it Adv. Colloid Interface Sci.} {\bf 70}, 125 (1997).
\bibitem{KMM49}
M.~Bordag, G.~L.~Klimchitskaya and
V.\ M.\ Mostepanenko,
{\it Phys. Rev. B} {\bf 86}, 165429 (2012).
\bibitem{KMM50}
M.~Chaichian, G.~L.~Klimchitskaya, V.\ M.\ Mostepanenko
and A.~Tureanu,
{\it Phys. Rev. A} {\bf 86}, 012515 (2012).
\bibitem{KMM51}
B.~Arora, H.~Kaur and B.~K.~Sahoo,
{\it J. Phys. B} {\bf 47}, 155002 (2014).
\bibitem{KMM52}
K.~Kaur, J.~Kaur, B.~Arora and B.~K.~Sahoo,
{\it Phys. Rev. B} {\bf 90}, 245405 (2014).
\bibitem{KMM53}
G.~L.~Klimchitskaya and V.~M.~Mostepanenko,
{\it Phys. Rev. A} {\bf 89}, 012516 (2014).
\bibitem{KMM54}
G.~L.~Klimchitskaya and V.~M.~Mostepanenko,
{\it Phys. Rev. B} {\bf 89}, 035407 (2014).
\bibitem{KMM55}
G.~L.~Klimchitskaya and V.~M.~Mostepanenko,
{\it Phys. Rev. A} {\bf 89}, 062508 (2014).
\bibitem{KMM56}
G.~L.~Klimchitskaya,
{\it Int. J. Mod. Phys. A} {\bf 31}, 1641026 (2016).
\bibitem{KMM57}
G.~Bimonte, G.~L.~Klimchitskaya and V.~M.~Mostepanenko,
{\it Phys. Rev. A} {\bf 96}, 012517 (2017).
\bibitem{KMM58}
M.~Bordag, I.~Fialkovsky, D.~Vassilevich,
{\it Phys. Lett. A} {\bf 381}, 2439 (2017).
\bibitem{KMM59}
C.~Henkel, G.~L.~Klimchitskaya and V.~M.~Mostepanenko,
{\it Phys. Rev. A} {\bf 97}, 032504 (2018).
\bibitem{KMM60}
N.~Khusnutdinov and N.~Emelianova,
{\it Int. J. Mod. Phys. A} {\bf 34}, 1950008 (2019).
\bibitem{KMM60a}
N.~Khusnutdinov and N.~Emelianova,
{\it Universe} {\bf 7}, 70 (2021).
\bibitem{KMM61}
G.~L.~Klimchitskaya,
{\it Mod. Phys. Lett. A} {\bf 35}, 2040004 (2020).
\bibitem{KMM62}
X.~Li, C.~W.~Magnuson, A.~Venugopal, R.~M.~Tromp, J.~B.~Hannon,
E.~M.~Vogel, L.~Colombo and R.~S.~Ruoff,
{\it J. Am. Chem. Soc.} {\bf 133}, 2816 (2011).
\bibitem{KMM63}
C.-C.~Chang, A.~A.~Banishev, R.~Castillo-Garza,
G.~L.~Klimchitskaya, V.\ M.\ Mostepanenko and U.\ Mohideen,
{\it Phys. Rev. B} {\bf 85}, 165443 (2012).
\bibitem{KMM64}
A.~A.~Banishev, C.-C.~Chang,
G. L.~Klimchitskaya, V.~M.\ Mostepanenko and U.\ Mohideen,
{\it Phys. Rev. B} {\bf 85}, 195422 (2012)
\bibitem{KMM65}
F.~J.~Giessibl,
{\it Rev. Mod. Phys.} {\bf 75}, 949 (2003).
\bibitem{KMM66}
C.~D.~Fosco, F.~C.~Lombardo and F.\ D.\ Mazzitelli,
{\it Phys. Rev. D} {\bf 84}, 105031 (2011).
\bibitem{KMM67}
G.~Bimonte, T.~Emig, R.\ L.\ Jaffe and M.\ Kardar,
{\it Europhys. Lett.} {\bf 97}, 50001 (2012).
\bibitem{KMM68}
G.~Bimonte, T.~Emig and M.~Kardar,
{\it Appl. Phys. Lett.} {\bf 100}, 074110 (2012).
\bibitem{KMM69}
M.~Bordag, G.~L.~Klimchitskaya and V.~M.~Mostepanenko,
{\it Int. J. Mod. Phys. A} {\bf 10}, 2661 (1995).
\bibitem{KMM70}
P.~J.~van Zwol, G.~Palasantzas and J.~Th.~M.~De Hosson,
{\it Phys. Rev. B} {\bf 77}, 075412 (2008).
\bibitem{KMM70a}
W. Broer,  G. Palasantzas, J. Knoester and V. B. Svetovoy,
{\it Phys. Rev. B} {\bf 85}, 155410 (2012).
\bibitem{KMM71}
W.~E.~Beadle, J.~C.~C.~Tsai and R.~D.~Plummer (eds.),
{\it Quick Reference Manual for Silicon Integrated Circuit Technology}
(Wiley, New York, 1985).
\bibitem{KMM72}
P.~Dai, Y.~Zhang and M.~P.~Sarachik,
{\it Phys. Rev. Lett.} {\bf 66}, 1914 (1991).
\bibitem{KMM73}
G.~L.~Klimchitskaya and V.~M.~Mostepanenko,
{\it Phys. Rev. B} {\bf 91}, 045412 (2015).
\bibitem{KMM75}
Grolltex, San Diego, CA, http://www.grolltex.com
\bibitem{KMM76}
MSE Supplies, Tucson, AZ, http://www.msesupplies.com
\bibitem{KMM77}
Y.~Wang, Y.~Zheng, X.~Xu, E.~Dubuisson, Q.~Bao, J.~Lu and K.~Ping Loh,
{\it ACS Nano} {\bf 5}, 9927 (2011).
\bibitem{KMM78}
G.~Li, A.~Luican and E.~Y.~Andrei,
{\it Phys. Rev. Lett.} {\bf 102}, 176804 (2009).
\bibitem{KMM79}
A.~Das, S.~Pisana, B.~Chakraborty, S.~Piscanec, S.~K.~Saha, U.~V.~Waghmare,
K.~S.~Novoselov, H.~R.\ Krishnamurthy, A.~K.~Geim, A.~C.~Ferrari and
A.~K.~Sood,
{\it Nature Nanotech.} {\bf 3}, 210 (2008).
\bibitem{KMM80}
L.~A.~Falkovsky,
{\it J. Phys.: Conf. Series} {\bf 129}, 012004 (2008).
\bibitem{KMM81}
L.~A.~Falkovsky,
{\it Pis'ma v ZETF} {\bf 98}, 183 (2013) [{\it JETP Letters} {\bf 98}, 161 (2013)].
\bibitem{KMM82}
G.~Bimonte,
{\it Europhys. Lett.} {\bf 118}, 20002 (2017).
\bibitem{KMM83}
M.~Hartmann, G.-L.~Ingold and P.~A.~Maia Neto,
{\it Phys. Rev. Lett.} {\bf 119}, 043901 (2017).
\bibitem{KMM84}
B.~Spreng, M.~Hartmann, V.~Henning, P.~A.~Maia Neto and G.-L.~Ingold,
{\it Phys. Rev. A} {\bf 97}, 062504 (2018).
\bibitem{KMM85}
M.~Hartmann, G.-L.~Ingold and P.~A.~Maia Neto,
{\it Phys. Scr.} {\bf 93}, 114003 (2018).
\bibitem{KMM86}
V.~Henning, B.~Spreng, M.~Hartmann, G.-L.~Ingold and P.~A.~Maia Neto,
{\it J. Opt. Soc. Amer. B} {\bf 36}, C77 (2019).
\bibitem{KMM87}
G.~L.~Klimchitskaya, V.~M.~Mostepanenko and V.~M.~Petrov,
{\it Phys. Rev. B} {\bf 96}, 235432 (2017).
\bibitem{KMM88}
G.~L.~Klimchitskaya, V.~M.~Mostepanenko and V.~M.~Petrov,
{\it Phys. Rev. A} {\bf 98}, 023809 (2018).
\bibitem{KMM89}
G.~L.~Klimchitskaya and V.~M.~Mostepanenko,
{\it Eur. Phys. J. C} {\bf 80}, 900 (2020).
\bibitem{KMM90}
G.~L.~Klimchitskaya and V.~M.~Mostepanenko,
{\it Phys. Rev. D} {\bf 103}, 096007 (2021).
\bibitem{KMM91}
G.~L.~Klimchitskaya and V.~M.~Mostepanenko,
{\it Phys. Rev. D} {\bf 104}, 085001 (2021).
\bibitem{KMM92}
G.~L.~Klimchitskaya and V.~M.~Mostepanenko,
{\it Phys. Rev. A} {\bf 105}, 012805 (2022).
\bibitem{KMM93}
M.~Bordag, I.~V.~Fialkovsky, N.\ Khusnutdinov and
D.~V.~Vassilevich,
{\it Phys. Rev. B} {\bf 104}, 195431 (2021).
\end{thebibliography}
\end{document}